# Physical-layer key distribution using synchronous complex dynamics of DBR semiconductor lasers

Anbang Wang[1,2,3,*], Yicheng Du[4,5], Qingtian Li[4,5], Longsheng Wang[4,5], Zhiwei Jia[4,5], Yuwen Qin[1,2,3], Yuncai Wang[1,2,3,*]

[1]*Institute of Advanced Photonics Technology, School of Information Engineering, Guangdong University of Technology, Guangzhou, 51006, China*

[2]*Key Laboratory of Photonic Technology for Integrated Sensing and Communication, Ministry of Education of China, Guangdong University of Technology, Guangzhou, 51006, China*

[3]*Guangdong Provincial Key Laboratory of Information Photonics Technology, Guangdong University of Technology, Guangzhou, 51006, China*

[4]*Key Lab of Advanced Transducers and Intelligent Control System, Ministry of Education, Taiyuan University of Technology, Taiyuan 030024, China*

[5]*College of Optoelectronics, Taiyuan University of Technology, Taiyuan 030024, China*

[*]*Corresponding author: abwang@gdut.edu.cn, wangyc@gdut.edu.cn*

**Abstract**

Common-signal-induced synchronization of semiconductor lasers with optical feedback inspired a promising physical key distribution with information-theoretic security and potential in high rate. A significant challenge is the requirement to shorten the synchronization recovery time for increasing key rate without sacrificing operation parameter space for security. Here, open-loop synchronization of wavelength-tunable multi-section distributed Bragg reflector (DBR) lasers is proposed as a solution for physical-layer key distribution. Experiments show that the synchronization is sensitive to two operation parameters, i.e., currents of grating section and phase section. Furthermore, fast wavelength-shift keying synchronization can be achieved by direct modulation on one of the two currents. The synchronization recovery time is shortened by one order of magnitude compared to close-loop synchronization. An experimental implementation is demonstrated with a final key rate of 5.98 Mbit/s over 160 km optical fiber distance. It is thus believed that fast-tunable multi-section semiconductor lasers opens a new avenue of high-rate physical-layer key distribution using laser synchronization.

# 1. Introduction

The increase of optical communication speed and the expansion of application scenarios have put forward urgent requirements for high-speed secure transmission techniques, especially which have compatibility with fiber communication link. Physical-layer key distribution that generates shared keys by physical principle is of vital importance to secure data transmission as it provides hardware-based security beyond algorithm encryption. Quantum key distribution is a typical secure physical-layer method but still has practical challenges in terms of speed and compatibility [1, 2]. Thus, key distribution based on classical optics has been receiving research attention, and methods utilizing giant fiber lasers [3-5], reciprocal channel noise [6-7], or synchronization of nonlinear oscillators [8-11] have been reported.

The key distribution method based on the common-signal-induced synchronization of optical nonlinear oscillators has attracted rising attention [10-21], since first demonstrated by Yoshimura *et al* [10]. In principle, a pair of nonlinear oscillators (such as semiconductor lasers) under injection of a common complex optical signal can emit synchronous irregular waveforms as physical entropy sources for generating shared keys. Due to that the synchronization is sensitive to device parameter mismatch, this method has information-theoretic security. In other words, it is practically difficult to eavesdrop full information of nonlinear oscillators without getting the same hardware device and knowing operation parameters in advance [10-13]. Recently, wideband chaos synchronization of semiconductor lasers over 1040 km fiber transmission has been experimentally achieved [22]. This means that the method has a potential comprehensive advantage of high key rate, long distance, and compatibility with fiber communication link.

Nonlinear oscillators based on semiconductor laser are the main candidate of the key distribution because of high complexity and easy-to-integrate configuration, which have been extensively used in physical random bit generation [23-25]. So far, semiconductor lasers including distributed-feedback (DFB) lasers [10, 13-17], vertical-cavity-surface emitting lasers (VCSELs) [18, 19], and Fabry-Perot lasers [11], have been demonstrated for key distribution experimentally or numerically. The reported approaches can be classified into two schemes. One is close-loop synchronization scheme using lasers with external optical feedback as entropy sources [10, 13-15, 18]. The feedback phase is a sensitive parameter for laser synchronization, and this enables random phase-shift-keying modulation by legitimate users to change laser state for enhancing security [10]. Unfortunately, the optical-feedback lasers have a long duration time of synchronization recovery (i.e., state-switching transient) in the magnitude of tens nanoseconds, and thus the key rate is greatly limited [13, 14]. For example, a key rate of 182 kb/s was experimentally obtained with optical-feedback DFB lasers [13]. Note that, the recovery time cannot be reduced by shortening the feedback cavity length even to a size of photonic integration [14]. The other one is open-loop synchronization scheme using lasers without external feedback [11, 16, 17, 19]. It shortens the synchronization recovery time but sacrifices sensitive laser parameters. As a result, external modulation for lasers is required as compensation of security, and some shift-keying modulation ways based on mode selection [11], dispersion [16], polarization [17], and time delay of self-carrier modulation [19] were presented. This leads to complicated system configuration and high cost.

In this paper, we propose using wavelength-tunable multi-section distributed Bragg reflector (DBR) semiconductor lasers as nonlinear oscillators for high-speed and secure key distribution with open-loop synchronization scheme. Experiments achieved wavelength-dependent synchronization of DBR lasers under common-signal injection, and revealed synchronization sensitivity to currents of grating section and phase section. This means that the DBR lasers themselves have sensitive parameters which can be used for user control by shift-keying modulation. Further, a short synchronization recovery time of several nanoseconds was obtained. Thus, the proposed method can not only increase key distribution rate but also keep security strength similar to close-loop synchronization scheme of DFB or VCSEL lasers. A final key rate of 5.98 Mbit/s was experimentally achieved over 160 km single-mode fiber link.

## 2. Experimental results

### 2.1. Principle

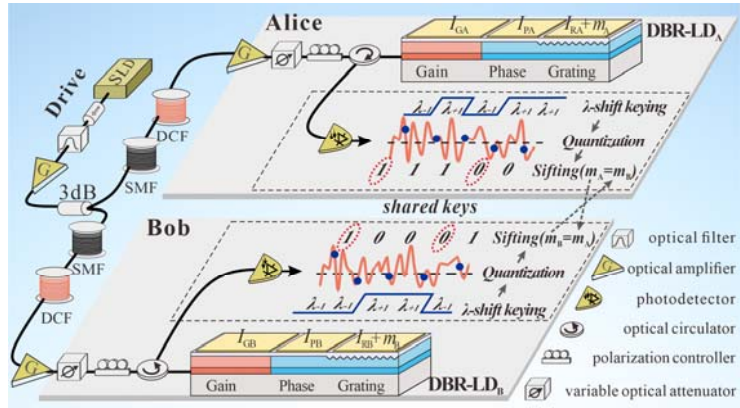

Fig.1 Schematic of experimental setup of key distribution using DBR-laser synchronization induced by a common complex signal. Wavelength-shift keying of lasers is realized by modulating currents $I_{RA,RB}$ (or $I_{PA,PB}$) using random and independent binary sequences $m_{A,B}$ and then random bits quantized from synchronous laser outputs (as $m_A=m_B$) are sifted as shared keys. SLD, super luminescent diode; SMF, single mode fiber; DCF, dispersion compensation fiber.

The schematic of experimental setup is shown in Fig. 1. A pair parameter-matched three-section DBR lasers denoted as DBR-LD$_A$ and DBR-LD$_B$ are allocated to legitimate users Alice and Bob, respectively. The lasers consist of a gain section, a phase section, and a Bragg grating section. The grating section currents are modulated as $I_{RA, RB} + m_{A, B}(t)$ to realize wavelength-shift keying between $\lambda_{-1}$ for $m_{A, B} < 0$ and $\lambda_{+1}$ for $m_{A, B} > 0$. Note, $I_{RA, RB}$ are bias currents and $m_{A, B}$ are two uncorrelated random bipolar non-return-zero (NRZ) sequences with a rate of $f_m$. The modulation can also be applied on phase section currents $I_{PA, PB}$. The drive signal is a spectrum-sliced spontaneous emission of a super luminescent diode (SLD); it is split by a 3dB coupler into two beams which are separately injected into the lasers after fiber transmission. The optical spectral width of drive light should cover both $\lambda_{-1}$ and $\lambda_{+1}$. The injection power of drive light is adjusted by a variable optical attenuator, and the polarization state is matched to the laser by a polarization controller. By precisely adjusting the injection power and the currents of gain and phase sections, the DBR lasers can achieve wavelength-shift keying synchronization. This means that synchronization randomly occurs during time slots as $m_A = m_B$. Next, the complex output waveforms of local DBR lasers are quantized into raw random binary sequences at a sampling frequency equal to $f_m$. Finally, users exchange sequences $m_A$ and $m_B$ in public and then sift shared keys in local from their own raw random numbers according to synchronization condition $m_A = m_B$.

Information-theoretic security based on bounded observability [10, 12] can be achieved under the following conditions. First, the correlation between the local laser output and the drive light is lower than that of two synchronized local lasers. Second, synchronization is sensitive to laser parameter mismatch so that it is practically hard for eavesdropper Eve to obtain well-matched lasers. As a result, Eve cannot observe the full information of the legitimate laser output. In addition, the keying modulation of laser state leads to that Eve has to get more lasers with the same number as that of the keying states [10]. Thus, using DBR lasers with direct modulation to replace DFB or VCSEL lasers can improve security strength of the open-loop-synchronization key distribution.

**2.2. Synchronization of DBR lasers**

In experiments, each fiber link between the drive and the DBR lasers consisted of a 66-km standard single mode fiber and a 14-km dispersion compensation fiber. The total transmission distance between the users was about 160km. Note that the drive source can also be placed at user end, with no influences on synchronization [22]. The DBR lasers (LD-PD Inc, PL-DBR-1550) with a threshold of $I_{th}$=37.8mA were stabilized by temperature controllers (ILX Lightwave LDT-5412), and the laser currents were controlled with a precision of 0.01mA (ILX Lightwave LDX-3210). An optical spectra analyzer (Finisar Wave Analyzer 1500S) with 1.2-pm resolution was used to monitor the wavelength matching. The NRZ modulation sequences were generated from an arbitrary waveform generator. The laser output waveforms were converted into electrical signals by photodetectors with 50-GHz bandwidth (Finisar XPDV2120R), and then sampled and quantized by a digital real-time oscilloscope (Lecroy LABMASTER10Z) with a bandwidth of 36 GHz.

We first investigated the wavelength-dependent synchronization of the DBR lasers with the following settings. The phase section currents were $I_{PA} = I_{PB} = 0.0$ mA. The gain section currents were $I_{GA} = 1.19 I_{th}$ and $I_{GB} = 1.17 I_{th}$ to ensure the lasers have the same relaxation frequency (2.1 GHz) which is one of the conditions of synchronization. The static-state wavelength of DBR-LD$_B$ was fixed at $\lambda_{-1}$ =1548.958 nm with a power of 412.2 µW by setting grating current $I_{RB}$ =12.10 mA. For DBR-LD$_A$, setting the grating current $I_{RA}$ = 13.41 mA yields the same wavelength $\lambda_{-1}$ with a power of 625.6 µW. Increasing $I_{RA}$ to 17.46 mA can tune the laser wavelength to 1548.429 nm (denoted as $\lambda_{+1}$) with a power of 242.1 µW. The drive light was adjusted to have a −3-dB spectral width of 0.827 nm with a center wavelength of 1548.840 nm to cover $\lambda_{-1}$ and $\lambda_{+1}$, which will be red-shifted due to optical injection. The injection powers into DBR-LD$_A$ and DBR-LD$_B$ were adjusted to 566 µW and 500 µW, respectively.

Figure 2 shows typical experimental results of wavelength-dependent synchronization. In the case of $\lambda_A = \lambda_B = \lambda_{-1}$, the lasers have identical static-state optical spectra depicted in dashed lines in Fig. 2(a). They have the same linewidth (−40-dB linewidth of 0.04 nm) and the same side modes with a spacing of 0.39 nm corresponding to the equivalent cavity length of DBR laser. After injection of the drive light, the side mode covered within the drive spectrum is gained greatly, and simultaneously is broadened and red-shifted together with the main mode. Shown in the solid lines, the broadened spectra are still identical. This indicates that synchronization is achieved. Fig. 2(b) presents a clear linear correlation between temporal waveforms, and the correlation coefficient is calculated as 0.97.

As tuning $\lambda_A$ from $\lambda_{-1}$ to $\lambda_{+1}$, desynchronization occurs. Shown in Fig. 2(c), the optical spectra of the two lasers are apparently different from each other. The scatterplot in Fig. 2(d) does not show correlation, and the corresponding correlation coefficient is only 0.11. This means that the information of the legitimate lasers cannot be fully obtained by an eavesdropper without knowing the wavelength setting beforehand, even if she were able to get a matched-well laser.

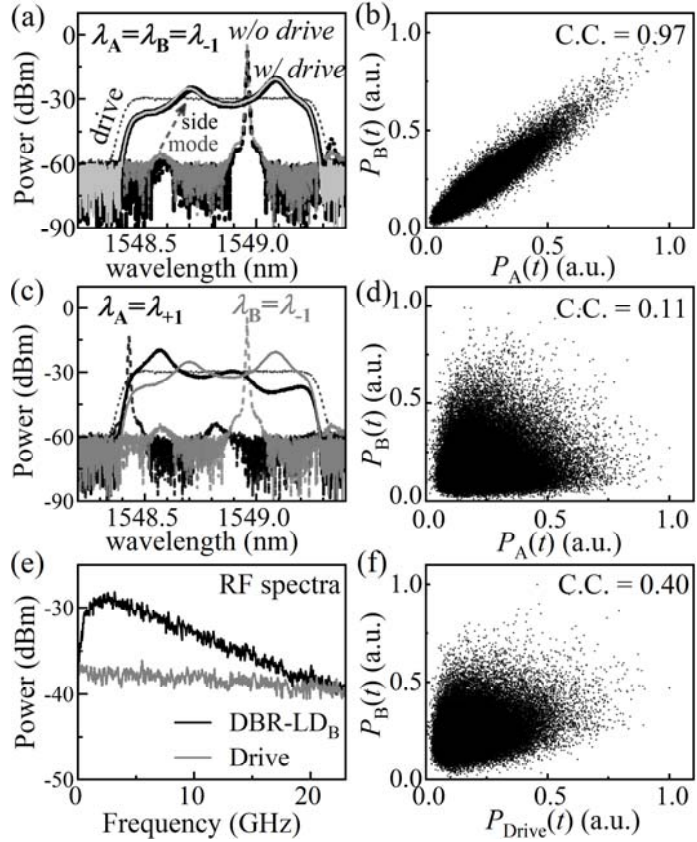

Fig. 2. Experimental results of wavelength-dependent synchronization: (a), (b) Optical spectra and waveform scatterplot of lasers with initial wavelength $\lambda_A = \lambda_B = \lambda_{-1} = 1548.958$nm; (c), (d) Optical spectra and waveform scatterplot of lasers with $\lambda_A = \lambda_{+1} = 1548.429$nm and $\lambda_B = \lambda_{-1}$; (e) electrical spectra and (f) correlation scatterplots of the laser and the drive light. C.C., correlation coefficient.

In addition, Figs. 2(e) and 2(f) show that the DBR laser response is also different from the drive signal. Obviously, their electrical spectra are significantly different in the frequency band of 0-15GHz. The profile of laser spectrum has an obvious peak locating at laser relaxation oscillation, while the drive spectrum is flat. The scatterplot of their time series in Fig. 2(f) does not display correlation. The correlation coefficient is calculated as about 0.4. This means that the information of laser output also cannot be fully eavesdropped by directly observing the drive signal. In brief, the wavelength-dependent synchronization and low correlation between the drive and the laser response guarantee the information-theoretic security based on bounded observability [10,12].

We further show the wavelength sensitivity of laser synchronization. Figs. 3(a) and 3(b) plot effects of the currents of grating and phase sections on synchronization (circles) and static-wavelength mismatch $\Delta\lambda = \lambda_A - \lambda_B$ (triangles) by fixing the parameters of DBR-LD$_B$, respectively. Seen from Fig. 3(a), the correlation coefficient reaches maximum as $\Delta\lambda=0$. Correlation coefficients beyond 0.9 is achieved as $I_{RA}$ changes within a range of −0.40 mA ~ 0.30 mA around the maximum. The corresponding wavelength mismatch is only within 0.01nm. In addition, as increasing $I_{RA}$ to about 16 mA, wavelength $\lambda_A$ jumps but still are covered by the drive spectrum, and correlation coefficients are reduced below 0.2. As decreasing $I_{RA}$ to about 10 mA, wavelength $\lambda_A$ jumps out of the drive spectrum, and the correlation coefficient tends to about 0.4. Shown in Fig. 3(b), phase section current $I_{PA}$ can also tune the laser wavelength to adjust the correlation coefficient of the two lasers. Correlation coefficients higher than 0.9 is achieved as $I_{PA}$ increases from 0 mA to 0.02 mA. This means synchronization is more sensitive to phase section current.

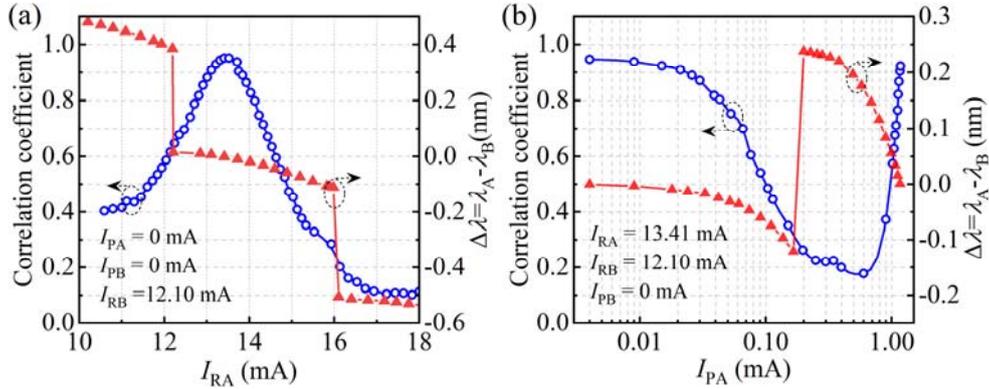

Fig. 3 Correlation coefficient and initial wavelength mismatch $\Delta\lambda = \lambda_A - \lambda_B$ as function of (a) DBR section current $I_{RA}$ and (b) phase section current $I_{PA}$.

Next, we demonstrate a typical experimental result of wavelength-shift keying synchronization between the DBR lasers by modulating grating section currents. We chose $\lambda_{-1}$ = 1548.958 nm and $\lambda_{+1}$ = 1548.429 nm as the two states of wavelength-shift keying modulation by setting $I_{RA}$ = 15.45 mA, $|m_A(t)|$ = 2.04 mA, $I_{RB}$ = 14.50 mA, and $|m_B(t)|$ = 2.40 mA. The modulation rate is $f_m$ = 80Mbit/s.

Figure 4(a) plots the random modulation sequences $m_A$ and $m_B$ in gray, the corresponding laser intensity waveforms $P_A$ and $P_B$ in blue and their short-term cross-correlation with a data length of 2 ns. Obviously, random on/off switching of laser synchronization is achieved; the synchronization occurs during time slots of $m_A=m_B$, otherwise desynchronization appears. The inset displays zoomed-in waveforms within a short time scale of 10ns to show laser synchronization. Shown in Fig. 4(b), a transient time of synchronization recovery of about 6.2 ns is measured. The histogram in Fig. 4(c) shows the mean value and standard deviation of the recovery time is about 5.4 ns and 1.5 ns, respectively. The synchronization recovery time is smaller than that of the photonic-integrate optical-feedback DFB lasers by one order of magnitude [14]. Note that, the laser waveforms within the transient of synchronization recovery are not synchronous and thus cannot be used to generate keys. Therefore, the reduction in the synchronization recovery time allows a higher modulation rate $f_m$ and then increase the key rate.

As shown in Fig. 4(d) the solitary DBR laser has a similar transient time of about 5 ns as switching lasing wavelength from $\lambda_{-1}$ to $\lambda_{+1}$. This indicates that the synchronization recovery time is determined by wavelength-switching induced transient time. The wavelength-switching time is mainly contributed by the time of carrier variation in the grating section until the non-lasing mode reaches threshold condition [27]. Therefore, optimizing laser parameters such as grating coupling coefficients could further reduce the transient time. In addition, modulation on phase section current also could reduce the recovery time, indicated by that synchronization is more sensitive to change of phase section current shown in Fig. 3.

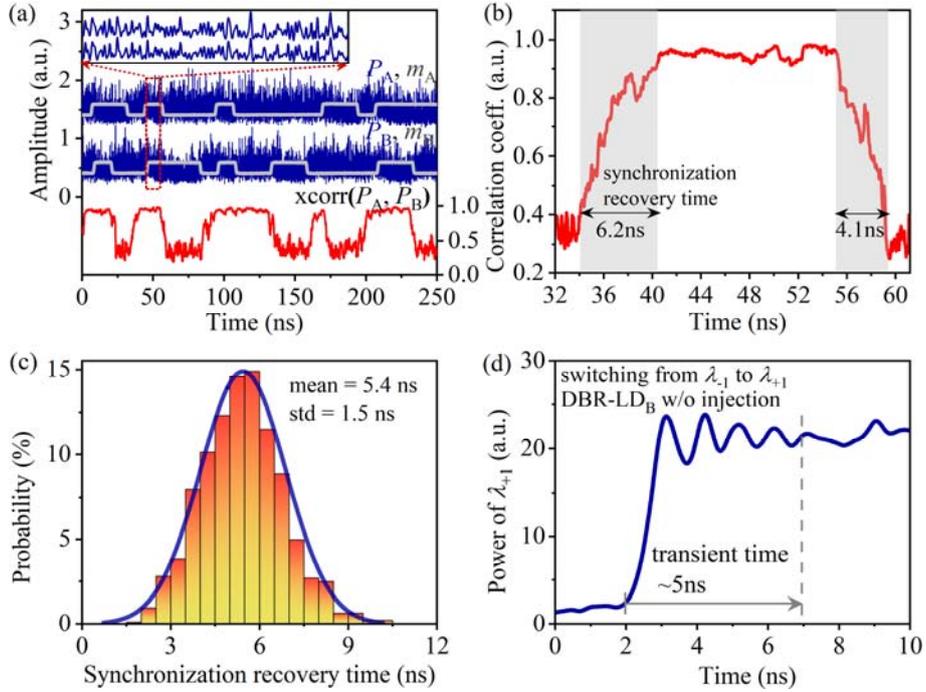

Fig. 4. Wavelength-shift keying synchronization: (a) temporal waveforms of laser outputs $P_{A, B}(t)$ and the corresponding keying codes $m_{A, B}(t)$, and the sliding short-term cross-correlation with a data length of 2 ns, and the inset shows zoomed-in synchronized waveforms in a 10-ns time window. (b) synchronization recovery time and (c) its probability distribution, (d) power transient of solitary laser under wavelength switching.

### 2.3. Key distribution

In the following, we analyze the key distribution using the synchronization of DBR lasers. First, the legitimate users independently extract raw random bits by dual-threshold quantization [10,11] with two voltage thresholds $V_u > V_m$ and $V_l < V_m$, where $V_m$ is the mean value of laser waveform. The sampling rate for random bit generation is equal to the modulation rate; that is, only one point is sampled per modulation period. Each sampling point is quantized as bit 1 or 0 if its voltage value is larger than $V_u$ or smaller than $V_l$, otherwise discarded. The points near the mean value which are susceptible to interference are discarded and then quantization robustness is enhanced. The two thresholds should be adjusted to make the extracted random bits unbiased. The number ratio of the extracted random bits to the sampling points is referred as to quantization retention ratio, denoted as $r_q$. Subsequently, users exchange and then compare their modulation codes and the indexes of the discarded bits to sift those random bits generated during the time slots of synchronization as the shared keys. The sifted-key rate is $R_{sift} = M^{-1} r_q f_m$, where $M = 2$ is the number of wavelength states and $M^{-1}$ is the probability of $m_A = m_B$. Note that, $R_{sift}$ can be estimated as the maximum key rate without considering error bits and information leakage.

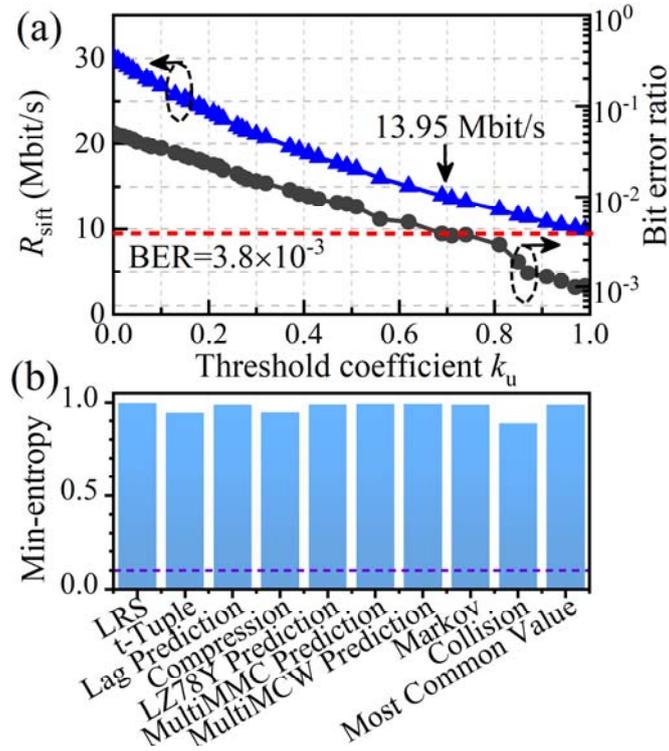

Fig. 5 (a) Sifted-key rate and BER as a function of comparison threshold $k_u$, (b) Min-entropy test results of laser output using NIST 800-90B.

Shown in Fig. 5(a), both the sifted-key rate and the corresponding BER decrease as increasing the normalized comparison threshold $k_u = (V_u - V_m)/\sigma$, where $\sigma$ is the standard deviation of laser waveform. As $k_u = 0.69$, a maximum key rate of 13.95 Mbit/s is achieved with a BER of $3.8 \times 10^{-3}$. The corresponding quantization retention ratio $r_q$ is equal to 0.35. Note that the BER value is a forward error correction (FEC) threshold BER, indicating that error bits in data stream caused by error keys under one-time-pad encryption can be corrected by FEC.

Let us estimate the final key rate by considering error bits and mutual information which could be observed by eavesdropper Eve. In general, the final key rate can be evaluated as $R_{final} = [1-\Sigma_{i=1}^{M}(I_{Ei} \cup I_D)/M - h(\text{BER})]R_{sift}$, where $I_{Ei}$ and $I_D$ represent information per bit known by Eve from the i[th] eavesdropping laser and from the drive signal, respectively. Assuming Eve has $M_E$ ($<M$) parameter-matched lasers with $I_{Ei}=1$ and $M-M_E$ mismatched lasers with the same $I_E$, one can obtain $R_{final} = [(1-M_E/M)(1-I_E-I_D)-h(\text{BER})]R_{sift}$ [10]. For simplicity, it is assumed that Eve uses the mean value for a single threshold to generate bits to evaluate $I_E$ and $I_D$ [13], and then $I_E=0.0014$ and $I_D=0.0675$ in our experiment. The entropy function $h(x) = -x\log_2(x) - (1-x)\log_2(1-x) = 0.0369$ for the FEC threshold BER of $3.8\times10^{-3}$. Due to that parameter-matched semiconductor lasers need to be selected from the same wafer [28], eavesdropper cannot obtain a matched laser if the manufacturer is trusted. In this case $M_E=0$ and the final key rate is roughly estimated as $R_{final}=12.47$ Mbit/s. Assuming a powerful eavesdropper who could get one matched laser ($M_E=1$), one can evaluate the final key rate as 5.98 Mbit/s. This final key rate is more than 32 times higher than that of the key distribution using photonic integrated optical-feedback DFB lasers [14].

Last, we examined the randomness of intensity waveform of the DBR laser by using NIST SP 800-90B test, which is the recommendation for entropy sources used for random bit generation [29]. This test includes ten items such as most common value, collision, and Markov, and if the min-entropy of all test items is larger than 0.1, the entropy source under test can be used to generate random bits. In our experiments, consecutive series with $10^6$ sampling points were used for test. The result in Fig.5(b) shows that the min-entropy is 0.89, which proves the randomness of the DBR laser output. Note the passing the randomness tests can be achieved as long as the laser wavelength is covered by the injection light spectrum.

## 3. Discussion and Conclusion

As demonstrated above, the common-signal-induced synchronization of DBR lasers has features such as low derive-response correlation and high parameter sensitivity, and thus satisfies the requirements of information-theoretic security based on bounded observability [12]. Especially, the DBR laser synchronization can be fast switched on-off by two operation parameters, i.e., currents of grating and phase sections. Compared to open-loop synchronization of DFB laser [26], it has stronger security because the DFB lasers do not have sensitive controllable parameters. Compared to closed-loop synchronization of optical-feedback DFB lasers [14], it not only has the same number of sensitive controllable parameters, but also has a shorter transient time. Therefore, the DBR-laser synchronization enables a faster secure key distribution. Designing lasers with more sections can further increase the security because the number of controllable parameters is increased. In addition, the distance of high-quality laser synchronization can be further extended through Raman and erbium-doped fiber hybrid amplification [22]. Therefore, the DBR-laser synchronization provides an alternative method of long-distance and high-speed physical-layer key distribution from a practical perspective.

It is noted that the output signals of the lasers remain private, and thus exchange of an additional information without breaking the overall privacy will be required to validate the laser synchronization. One feasible method is using a narrow spectrum-sliced component of the lasers as the additional exchange information, and using the remaining part as entropy source to generate random numbers in private. The synchronization of the exchanged spectrum-sliced signals means the laser synchronization is built. in addition, the exchanged signal has no correlation to the entropy signal, and thus the privacy is not broken.

To summarize, the physical-layer key distribution using common-noise-induced synchronization of three-section DBR semiconductor lasers with wavelength-shift keying modulation is proposed and experimentally demonstrated. Experimental results show that the synchronization is sensitive to laser wavelength mismatch with a small tolerance of about 0.01nm, and direct modulation on currents of grating section or phase section can realize wavelength-shift keying synchronization. A short average synchronization recovery time of about 5.4 ns was obtained. Therefore, the key distribution using the DBR lasers can increase key rate without reduction of security strength, compared to optical-feedback single mode lasers. Typically, a final key distribution rate of 5.98 Mbit/s over 160 km optical fiber distance is achieved experimentally. This work pays a way for high-rate physical key distribution method with information-theoretic security by using fast-tunable multi-section semiconductor lasers.

**Acknowledgements**

This work was partially supported by the National Natural Science Foundation of China (62035009, U22A2087, 61731014, 61822509), and the Program for Guangdong Introducing Innovative and Enterpreneurial Teams.